\documentclass[letterpaper]{article}
\usepackage{aaai2026}
\usepackage{times}
\usepackage{helvet}
\usepackage{courier}
\usepackage[hyphens]{url}
\usepackage{graphicx}
\usepackage{booktabs}
\usepackage{enumitem}

\usepackage{natbib}
\usepackage{caption}
\title{Understanding Student Attitudes and Acceptability of GenAI Tools in Higher Ed: Scale Development and Evaluation}
\author {
    Xiuxiu Tang,
    Si Chen,
    Ying Cheng,
    Nitesh V Chawla,
    Ronald Metoyer,
    G. Alex Ambrose
}
\affiliations {
    University of Notre Dame\\
    \{xtang8, schen34, ycheng4, nchawla, rmetoyer, gambrose\}@nd.edu
}

\begin{document}

\maketitle

\begin{abstract}
As generative AI (GenAI) tools like ChatGPT become more common in higher education, understanding student attitudes is essential for evaluating their educational impact and supporting responsible AI integration. This study introduces a validated survey instrument designed to assess students’ perceptions of GenAI, including its acceptability for academic tasks, perceived influence on learning and careers, and broader societal concerns. We administered the survey to 297 undergraduates at a U.S. university. The instrument includes six thematic domains: institutional understanding, fairness and trust, academic and career influence, societal concerns, and GenAI use in writing and coursework. Exploratory factor analysis revealed four attitudinal dimensions: societal concern, policy clarity, fairness and trust, and career impact. Subgroup analyses identified statistically significant differences across student backgrounds. Male students and those speaking a language other than English at home rated GenAI use in writing tasks as more acceptable. First-year students expressed greater societal concern than upper-year peers. Students from multilingual households perceived greater clarity in institutional policy, while first-generation students reported a stronger belief in GenAI’s impact on future careers. This work contributes a practical scale for evaluating the student impact of GenAI tools, informing the design of educational AI systems.
\end{abstract}

\section{Introduction}

Generative artificial intelligence (GenAI) tools—such as ChatGPT—are rapidly transforming higher education by reshaping how students engage with academic writing, coding, idea generation, and exam preparation. These tools offer significant promise for learning efficiency and academic support. However, their widespread adoption has also raised pressing concerns around academic integrity, digital equity, and institutional policy clarity~\cite{cao2025students,johnston2024student,baek2024chatgpt,chan2023students}. As GenAI becomes increasingly embedded in educational practice, there is an urgent need for validated, student-centered measures to evaluate its real-world impact and support inclusive, responsible AI integration. 

Despite growing interest, key stakeholders—including technology developers, researchers, policymakers, instructors, and institutional leaders—currently lack robust, psychometrically sound instruments to assess students’ attitudes toward GenAI tools.  Early studies have begun documenting student experiences and concerns, but many rely on ad hoc or custom survey items, lack construct validity, or fail to capture the multidimensional nature of GenAI attitudes. Moreover, few systematically examine how perceptions vary across demographic subgroups—such as gender, first-generation college status, or home language—limiting our understanding of equity in AI adoption. Existing validated instruments remain rare in U.S. undergraduate contexts and are seldom used to explore subgroup variation~\cite{marengo2025development,mogelvang2025validating}.

To address these gaps, this paper contributes a \textbf{validated survey instrument} designed to collect and measure students’ attitudes and perceived acceptability of GenAI tools in higher education. We present findings from a field study involving over 297 undergraduate students at U.S., mid-sized, highly-ranked private research university, evaluating the instrument’s reliability and exploring how attitudes vary across demographic groups. By combining rigorous survey design with real-world evaluation, our work directly responds to calls for socially responsible AI research through improved data collection, bias mitigation, and impact assessment in educational settings. We answer the following research questions: \textbf{RQ1}:
How consistent and reliable are the survey subscales designed to measure students’ attitudes and perceived acceptability of Gen AI tools in educational settings? \textbf{RQ2}:
What are students’ attitudes and perceived acceptability toward Generative AI tools, and how do these vary by demographic factors?

\section{Related Work}

\subsection{Student Use of GenAI in Higher Education}

The rapid integration of GenAI tools in higher education has prompted a wave of studies documenting how students use these technologies for academic purposes. Across global contexts, students report using GenAI for writing assistance, grammar correction, brainstorming, coding support, and summarization tasks \cite{ravvselj2025higher, cao2025students, baidoo2024exploring, mogelvang2025validating, marengo2025development}. These tools are often perceived as extensions of students’ learning strategies, especially in writing-intensive or conceptually demanding coursework.

While many students report benefits from using GenAI tools, such as improved efficiency and writing support, studies have also documented student concerns about academic integrity, overreliance on GenAI, diminished critical thinking, fairness in evaluation processes, and risks related to data privacy and security \cite{maxwell2025generative, krvavica2025students, johri2024misconceptions}. These concerns are often compounded by uncertainty stemming from unclear or inconsistent institutional policies, particularly regarding assessment and acceptable use practices.

Recent evidence suggests that students’ attitudes toward ChatGPT use are shaped not only by perceived benefits and risks but also by their usage practices. \citet{acosta2024analysis} found that students who used ChatGPT responsibly and verified its output expressed more positive attitudes across dimensions such as ethical concern, learning support, and intention of use. In contrast, those who did not verify information or used the tool irresponsibly were more likely to express negative perceptions. To support meaningful and ethical academic engagement, it is important not only to provide access to GenAI tools but also to cultivate responsible usage habits and critical digital literacy among students.

\subsection{Demographic Variation in Attitudes}

Recent studies suggest that attitudes toward GenAI vary across demographic groups, though findings remain inconsistent. Male students and STEM majors often report more favorable perceptions of GenAI and greater confidence in its use, whereas female students and those in non-STEM disciplines are more likely to express ethical concerns and uncertainty about appropriate use \cite{baek2024chatgpt, mogelvang2025validating, sublime2024chatgpt}. However, some studies report contrasting trends. For example, \citet{sun2024generative} found that female students and those with prior experience using GenAI held more positive attitudes, with education majors scoring significantly higher than students in other disciplines.

\citet{mogelvang2025validating} similarly identified gender, age, and field of study as significant predictors of GenAI attitudes in a large sample of Norwegian university students. Male students and those aged 21–25 expressed more positive attitudes than female students and those younger than 21 or older than 25. Additionally, students from engineering and natural sciences were more positive than their peers in health sciences and teacher education, culture, and sports, though not significantly more positive than those in economics and social sciences.

Language background and institutional context also appear to influence GenAI engagement. \citet{baek2024chatgpt} reported that non-native English speakers were more likely than native speakers to use ChatGPT for writing tasks, suggesting its perceived value as a language support tool. Furthermore, students in their 30s and 40s reported higher usage than younger peers, and institutional policies permitting ChatGPT use were associated with greater adoption.

However, not all studies observe demographic effects. \citet{baidoo2024exploring} found no significant variation in students' perceptions of ChatGPT by gender, age, or education level, which suggests that students across backgrounds generally share similar views on its usefulness and accessibility. These mixed findings underscore the importance of context-sensitive and intersectional analyses that consider cultural, institutional, and technological factors.

\subsection{Limitations of Existing Measurement Approaches}

Although student perceptions of GenAI are receiving growing scholarly attention, much of the existing literature still relies on self-constructed or loosely validated survey instruments. Many studies employ descriptive statistics or qualitative analyses of open-ended responses, which limits the generalizability, replicability, and comparability of findings across educational contexts and student populations \cite{krvavica2025students, johri2024misconceptions, johnston2024student, baek2024chatgpt}. While these exploratory approaches offer initial insights, their methodological limitations hinder robust conclusions about the structure and variability of GenAI attitudes.

Only a limited number of studies have adopted validated psychometric tools to assess students’ attitudes toward GenAI. For example, \citet{marengo2025development} developed a 13-item GenAI Attitude Scale encompassing both positive and negative attitudes. Their scale demonstrated high internal consistency and construct validity, capturing diverse dimensions such as perceived usefulness, ease of use, learning impact, ethical concerns, and perceived barriers. Similarly, \citet{mogelvang2025validating} introduced the 4-item AI Attitude Scale (AIAS-4), a brief and adaptable instrument that measures general positive attitudes toward AI, also showing strong reliability and structural validity.

However, these instruments have notable limitations for use in U.S. higher education contexts. Most were developed in non-U.S. settings or with general public samples, raising questions about their applicability to American undergraduates. Moreover, few validated scales have been used to examine subgroup differences (e.g., by gender, age, language background, or major) or to connect students’ attitudes with institutional policies, pedagogical supports, or ethical concerns. To effectively capture the complex, domain-specific dimensions of GenAI attitudes in educational settings, there is a clear need for instruments that are both contextually grounded and psychometrically robust.

Building on this foundation, our study integrates psychometric validation and demographic analysis to explore the complex and varied landscape of GenAI perceptions in higher education. By doing so, we contribute a theoretically grounded and practically relevant perspective to ongoing discussions about the role of AI in learning environments.

\section{Methods}

\subsection{Participants}

A total of 297 undergraduate students participated in this study, all from a mid-sized, highly ranked private research university in the United States known for its strong undergraduate teaching. Recruitment occurred in Spring 2025 through institutional mailing lists and course announcements. All participants provided informed consent and completed the anonymous online survey via Qualtrics. The study protocol was approved by the university’s Institutional Review Board (IRB).

The sample reflected a diverse cross-section of the undergraduate population. In terms of gender identity, 55\% identified as women, 43\% as men, and 3\% selected another option. The majority of respondents (88\%) were domestic students, with 12\% identifying as international. Regarding academic background, 61\% of participants reported pursuing non-STEM majors, while 39\% were enrolled in STEM-related programs. Most students (59\%) were in their first year of study, and 41\% had been enrolled for two or more years. Additionally, 84\% primarily spoke English at home, and 16\% reported using a language other than English. Ten percent of the sample identified as first-generation college students, and 13\% self-reported having a disability, health or mental health condition, or being neurodivergent.

\subsection{Survey Instrument Development and Structure}

The survey instrument was developed through a multi-stage process grounded in literature on GenAI and higher education. An initial pool of items was created by three domain experts with backgrounds in educational measurement, instructional design, and AI in education. Item development was informed by key themes identified in prior research, including a previously published survey on GenAI perceptions in higher education \cite{chung2024student}. To ensure content validity and conceptual clarity, a panel of additional experts in higher education, ethics, and AI policy reviewed the draft items and provided feedback on item wording, coverage, and alignment with the intended constructs. Revisions were made accordingly to improve clarity, remove redundancy, and enhance relevance to undergraduate student experiences.

The final survey included six thematic sections, each aligned with important dimensions of student attitudes and use patterns. Items in Groups G1–G4 measured attitudinal beliefs using a 5-point Likert scale (1 = Strongly Disagree to 5 = Strongly Agree):

\begin{itemize}
\item \textbf{G1:} Perceptions of GenAI and institutional policy clarity
\item \textbf{G2:} Concerns about fairness, trust, and instructor judgment
\item \textbf{G3:} Perceived impact of GenAI on students’ educational and career trajectories
\item \textbf{G4:} Broader societal concerns, including bias, privacy, and environmental impact
\end{itemize}

Groups G5 and G6 assessed the perceived acceptability of GenAI use in specific academic tasks using a 5-point acceptability scale (1 = Always Unacceptable to 5 = Always Acceptable):

\begin{itemize}
\item \textbf{G5:} Acceptability of GenAI in writing-related activities (e.g., grammar correction, idea generation, paraphrasing)
\item \textbf{G6:} Acceptability of GenAI in coursework and study support (e.g., generating study guides, supporting coding or language translation)
\end{itemize}

The survey also collected self-reported demographic data, including gender identity, enrollment type (domestic/international), area of study, years enrolled, home language, first-generation college status, and disability or neurodivergence identification. See Appendix A for the full survey.

\subsection{Data Preparation}

The original dataset included responses from 444 students. To ensure data quality and minimize the impact of missingness, we excluded cases with more than 30\% missing values across the survey and demographic variables. This threshold retained 297 cases for analysis. Given the relatively low level of item-level missingness in retained cases, missing data were handled using listwise deletion within each analysis.

Demographic variables were recoded into binary categories to facilitate group comparisons (e.g., Man vs. Woman; STEM vs. Non-STEM; $\leq$1 year vs. $>$1 year). Responses marked as ``Other'' or ``Prefer not to say'' were excluded from inferential analyses due to small subgroup sizes. Reverse coding was applied to selected items in Groups G2 and G4 to ensure conceptual alignment, such that higher scores consistently reflected more favorable or more accepting attitudes toward GenAI.

\subsection{Analysis}

We analyzed the survey data in four stages. First, we summarized how students responded to individual survey items across the six thematic domains. Second, we assessed the reliability of each item group and refine the item groups based on reliability analyses. Third, we used exploratory factor analysis (EFA) to identify underlying themes in students’ attitudes. Finally, we examined whether these scores differed across demographic groups using statistical comparisons.

\subsubsection{Psychometric Analysis}

To evaluate internal consistency, we calculated Cronbach’s alpha for each item group. This statistic estimates how closely related a set of items are as a group, indicating whether they are measuring the same underlying concept. Higher alpha values (closer to 1) indicate stronger internal consistency. As a general guideline, values above 0.70 are considered acceptable, above 0.80 are considered good, and values above 0.90 are considered excellent for research purposes.

For Groups G1–G4, we conducted EFA to detect underlying attitudinal dimensions based on patterns in student responses. This analysis used the minimum residual (minres) extraction method and oblimin rotation, which allows the identified dimensions to be correlated. We checked the suitability of the data for EFA using the Kaiser-Meyer-Olkin (KMO) measure and Bartlett’s test of sphericity. The KMO statistic assesses whether the survey items share enough common variance to justify using factor analysis, with values closer to 1 indicating better suitability. Bartlett’s test checks whether the items are sufficiently correlated overall to support factor extraction.

To determine how many factors to retain, we used parallel analysis and scree plot inspection. Parallel analysis compares the observed data to randomly generated data to identify meaningful factors. The scree plot is a visual tool that helps detect the point at which additional factors add little explanatory value.

We then computed factor scores using the regression method, which creates a weighted summary score for each participant based on their responses and the factor loadings. These scores were used in subsequent analyses, including group comparisons.

Although Groups G5 and G6 were not included in the EFA, we evaluated their internal consistency. Both groups demonstrated strong reliability ($\alpha = 0.78$ for G5, $0.82$ for G6). Because these items were designed to measure focused concepts and showed sufficient consistency, we calculated each group’s score by averaging item responses. This approach preserved interpretability on the original 5-point scale and offered a simple, valid summary of each domain.

\subsubsection{Group Comparisons}

To examine whether students’ attitudes toward GenAI differed by demographic background, we compared average scores between groups using Welch’s \textit{t}-tests. This statistical method is used to test whether the means of two groups differ significantly and is particularly appropriate when the groups have unequal sample sizes or variances—a common feature in survey data.

We analyzed six outcome variables: four standardized factor scores derived from the attitudinal dimensions identified through EFA (G1–G4), and two composite scores (G5 and G6), which reflect students’ perceived acceptability of GenAI use in writing and coursework tasks.

To evaluate the size or practical importance of the observed group differences, we calculated Cohen’s \textit{d}, a widely used measure of effect size. It expresses the difference between two group means in standardized units. In social science research, values of $d = 0.2$, $0.5$, and $0.8$ are commonly interpreted as small, medium, and large effects, respectively \cite{cohen2013statistical}. These thresholds provide general guidance, though real-world effects in education and psychology are often smaller and should be interpreted in context. All statistical analyses were conducted using R.

\section{Results}

\subsection{Descriptive Patterns Across Thematic Groups}

\begin{figure}[ht]
    \centering
    \includegraphics[width=0.48\textwidth]{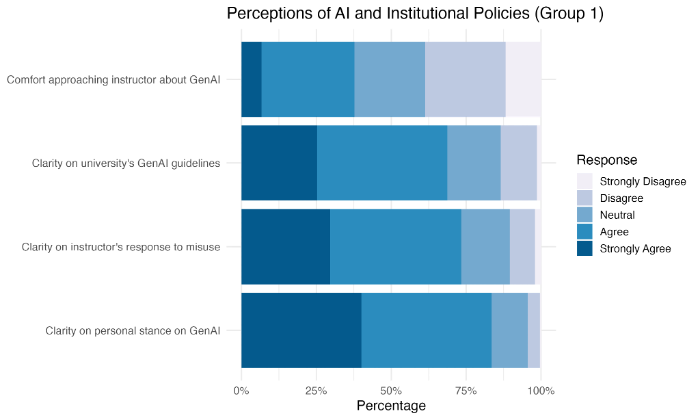}
    \includegraphics[width=0.48\textwidth]{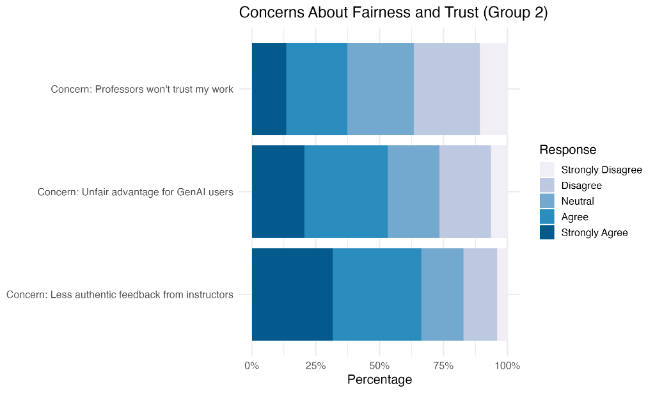}
    \caption{Item-level response distributions for Group 1 (Institutional Policies) and Group 2 (Fairness and Trust).}
    \label{fig:g1_g2}
\end{figure}

\begin{figure}[ht]
    \centering
    \includegraphics[width=0.48\textwidth]{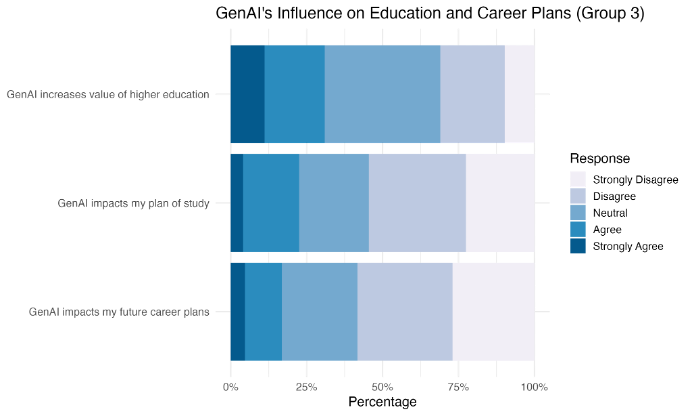}
    \includegraphics[width=0.48\textwidth]{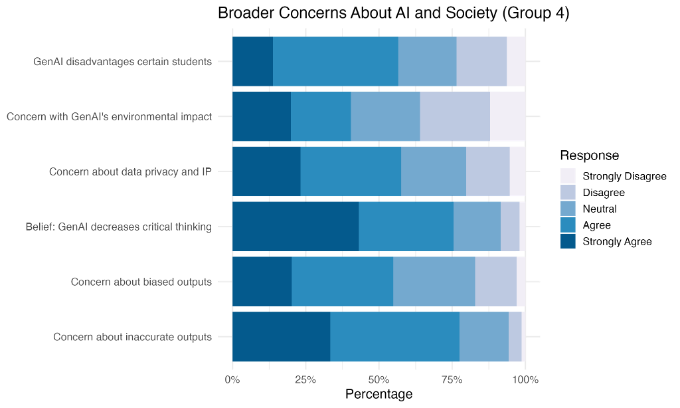}
    \caption{Item-level response distributions for Group 3 (Education and Career Impact) and Group 4 (Broader Societal Concerns).}
    \label{fig:g3_g4}
\end{figure}

\begin{figure}[ht]
    \centering
    \includegraphics[width=0.48\textwidth]{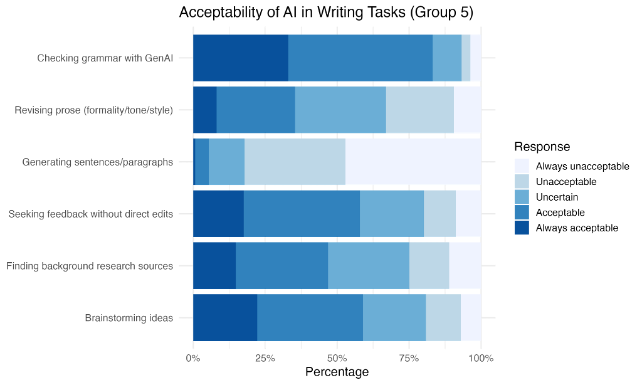}
    \includegraphics[width=0.48\textwidth]{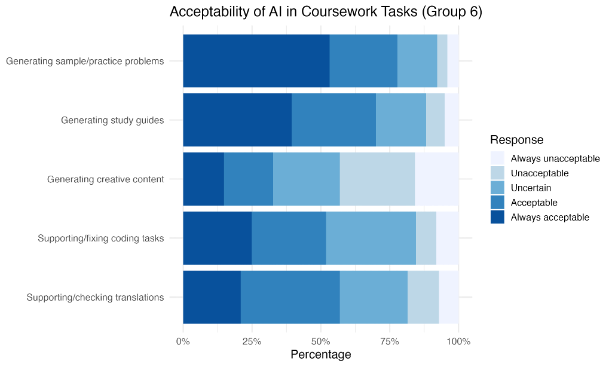}
    \caption{Item-level response distributions for Group 5 (Writing Task Acceptability) and Group 6 (Coursework Task Acceptability).}
    \label{fig:g5_g6}
\end{figure}

To establish an initial understanding of students’ attitudes toward GenAI, we examined the distribution of responses across the six conceptual domains captured in the survey. Figures~\ref{fig:g1_g2} through~\ref{fig:g5_g6} present stacked bar charts that visualize the percentage of students selecting each response category on the five-point Likert scale. These visualizations offer nuanced insight into attitudinal variation that may not be apparent from aggregate statistics alone.

Students’ perceptions of institutional policy (G1) were relatively mixed. While a subset of respondents agreed that instructors and institutions had communicated clear expectations regarding GenAI use, a notable portion expressed uncertainty or disagreement—suggesting variability in policy awareness and perceived transparency. This ambiguity is echoed in responses related to fairness and trust (G2), where many students voiced concerns about the equitable use of GenAI and whether instructors would fairly assess student work in the presence of AI-generated content.

In contrast, responses in the domain of educational and career influence (G3) indicated that students were less uniformly convinced of GenAI’s transformative impact on their academic or professional pathways. Although some acknowledged benefits, a significant number remained neutral or unconvinced, implying cautious optimism rather than wholesale endorsement. Meanwhile, broader societal concerns (G4) were more widely acknowledged. A majority of students recognized the ethical and social risks associated with GenAI, including potential bias, misinformation, and data privacy—reflecting a relatively high level of critical awareness.

When asked about the acceptability of GenAI use in specific academic tasks, clear differences emerged between writing and coding contexts. Students were generally more accepting of GenAI use in lower-stakes or process-oriented tasks, such as brainstorming ideas or debugging code (G5 and G6). However, opinions were more divided when it came to using GenAI to generate or revise academic prose, with some students expressing ethical reservations or discomfort.

These descriptive patterns reveal the complexity of student perspectives: while many recognize the utility of GenAI tools, concerns about fairness, institutional guidance, and long-term implications remain salient.

\subsection{Scale Reliability and Item Refinement}

We evaluated the internal consistency of each item group using Cronbach’s $\alpha$ and examined item-total correlations to identify items that weakened reliability or lacked conceptual alignment. Items with poor psychometric performance were excluded, resulting in refined scales with acceptable to strong reliability ($\alpha = 0.59$ to $0.82$). Table~\ref{tab:reliability_summary} summarizes the number of retained items and final $\alpha$ coefficients for each group. Table~\ref{tab:item_stats} presents item-total correlations, all of which exceeded the commonly accepted threshold of 0.30, supporting the internal consistency of the final scales.

The institutional policy scale (G1) initially showed moderate reliability ($\alpha = 0.67$). Removing item G1\_1 improved reliability to 0.71. The fairness and trust scale (G2) exhibited poor internal consistency at first ($\alpha = 0.48$); excluding item G2\_1 raised the $\alpha$ to 0.59. For the education and career impact scale (G3), reliability increased substantially—from 0.59 to 0.81—after removing item G3\_1, suggesting that it did not align conceptually with the rest of the scale.

The societal concerns scale (G4) demonstrated acceptable reliability ($\alpha = 0.76$), and all items were retained. The GenAI acceptability in writing tasks scale (G5) showed good reliability ($\alpha = 0.78$). The GenAI acceptability in coursework tasks acceptability scale (G6) showed the highest reliability ($\alpha = 0.82$), with no item modifications required.

\begin{table}[ht]
\centering
\small
\caption{Reliability Summary for Each Survey Scale}
\label{tab:reliability_summary}
\begin{tabular}{p{1cm}p{6cm}}
\toprule
Group & Construct and Reliability \\
\midrule
G1 & Policy Perception: 3 items; $\alpha = 0.71$ \\
G2 & Fairness/Trust Concerns: 3 items; $\alpha = 0.59$ \\
G3 & Education \& Career Impact: 2 items; $\alpha = 0.81$ \\
G4 & Broader Societal Concerns: 5 items; $\alpha = 0.76$ \\
G5 & Writing Task Acceptability: 6 items; $\alpha = 0.78$ \\
G6 & Coursework Task Acceptability: 5 items; $\alpha = 0.82$ \\
\bottomrule
\end{tabular}
\end{table}

\begin{table}[ht]
\centering
\small
\caption{Descriptive Statistics and Corrected Item-Total Correlations for Retained Survey Items}
\label{tab:item_stats}
\begin{tabular}{llccc}
\toprule
Group & Item & Mean & SD & Item-Total Corr. \\
\midrule
G1 & G1\_2 & 3.79 & 0.99 & 0.57 \\
   & G1\_3 & 3.91 & 0.99 & 0.62 \\
   & G1\_4 & 4.19 & 0.82 & 0.42 \\
\midrule
G2 & G2\_2 & 3.41 & 1.20 & 0.46 \\
   & G2\_3 & 3.77 & 1.15 & 0.35 \\
   & G2\_4 & 3.40 & 1.12 & 0.38 \\
\midrule
G3 & G3\_2 & 2.49 & 1.15 & 0.69 \\
   & G3\_3 & 2.36 & 1.14 & 0.69 \\
\midrule
G4 & G4\_1 & 3.12 & 1.31 & 0.41 \\
   & G4\_2 & 3.55 & 1.16 & 0.60 \\
   & G4\_3 & 4.08 & 1.01 & 0.49 \\
   & G4\_4 & 3.55 & 1.06 & 0.68 \\
   & G4\_5 & 4.04 & 0.89 & 0.52 \\
\midrule
G5 & G5\_1 & 4.06 & 0.94 & 0.58 \\
   & G5\_2 & 3.01 & 1.10 & 0.57 \\
   & G5\_3 & 1.77 & 0.89 & 0.42 \\
   & G5\_4 & 3.47 & 1.16 & 0.61 \\
   & G5\_5 & 3.26 & 1.20 & 0.43 \\
   & G5\_6 & 3.55 & 1.17 & 0.55 \\
\midrule
G6 & G6\_1 & 4.19 & 1.07 & 0.65 \\
   & G6\_2 & 3.93 & 1.14 & 0.65 \\
   & G6\_3 & 2.89 & 1.29 & 0.53 \\
   & G6\_4 & 3.53 & 1.18 & 0.60 \\
   & G6\_5 & 3.52 & 1.15 & 0.67 \\
\bottomrule
\end{tabular}
\vspace{1mm}
\begin{flushleft}
\footnotesize \textit{Note.} Items G1\_1, G2\_1, and G3\_1 were excluded during scale refinement and are not shown here.
\end{flushleft}
\end{table}

\subsection{Exploratory Factor Analysis (EFA)}

To examine the underlying structure of students’ attitudes toward GenAI, we conducted an EFA on 13 attitudinal items from Groups G1 through G4. Before extracting factors, we assessed the suitability of the data. The KMO measure of sampling adequacy was 0.73, exceeding the commonly recommended threshold of 0.60. Item-level measures of sampling adequacy values ranged from 0.54 to 0.88, indicating acceptable shared variance. Bartlett’s test of sphericity was significant, $\chi^2$(78) = 1087.83, $p < .001$, suggesting that the correlation matrix was appropriate for factor analysis.

We used parallel analysis to determine the appropriate number of factors to retain. The scree plot (Figure~\ref{fig:scree} in Appendix B) compares eigenvalues from the observed data (blue line) with those from randomly simulated (red dotted line) and resampled (red dashed line) datasets. The first four observed eigenvalues exceeded both reference lines, supporting a four-factor solution. Based on this result, we conducted exploratory factor analysis (EFA) using the minimum residual (minres) extraction method with oblimin rotation, which permits correlations among factors. The resulting four-factor model explained 47.3\% of the total variance, with individual factors accounting for 15.9\%, 11.5\%, 11.2\%, and 8.7\%, respectively. Model fit indices suggested good fit: root mean square residual (RMSR) = 0.03, Tucker-Lewis Index (TLI) = 0.934, and root mean square error of approximation (RMSEA) = 0.053 with 90\% CI [0.031, 0.075].

\begin{figure}[ht]
\centering
\includegraphics[width=0.4\textwidth]{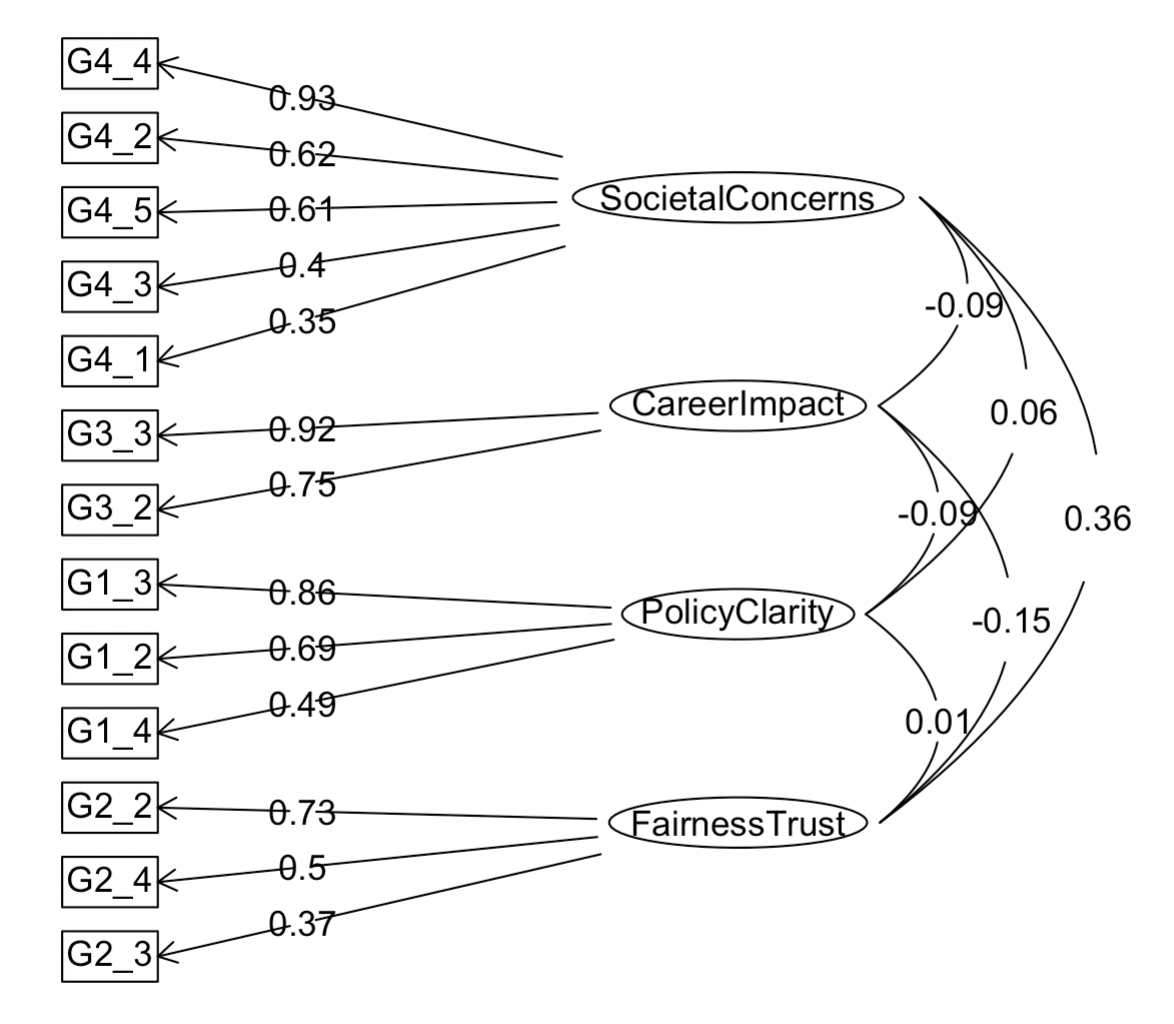}
\caption{Path diagram showing factor loadings and inter-factor correlations.}
\label{fig:path}
\end{figure}

Figure~\ref{fig:path} displays a path diagram of the factor structure. Each ellipse represents a latent factor, and arrows denote the primary loading for each item. The diagram highlights the conceptual coherence of the item clusters and visualizes the relationships among factors. Inter-factor correlations ranged from $r = -0.15$ to $r = 0.36$, suggesting some shared variance while also supporting their distinctiveness. 

The factor loadings revealed four interpretable dimensions. The first factor, \textit{Societal Concerns}, included all five items from Group G4 and captured students’ concerns about AI’s societal implications (e.g., bias, privacy, environmental impact). The second factor, \textit{Career Impact}, was defined by two items from G3 reflecting beliefs about GenAI’s relevance for academic and professional trajectories. The third factor, \textit{Policy Clarity}, comprised three items from G1 measuring perceived clarity of institutional rules. The fourth factor, \textit{Fairness and Trust}, reflected perceptions of fairness in GenAI policy and instructor judgment, based on three items from G2.

\subsection{Demographic Comparisons}

Several statistically significant differences in student attitudes toward GenAI were observed across demographic groups (see Table~\ref{tab:group_diffs}). Compared to female students, male students reported more favorable views of using GenAI for writing tasks ($d = 0.16$, $p = .012$), which represents a small effect size. Students who primarily spoke a language other than English at home also showed greater acceptance of GenAI for writing tasks than their English-speaking peers ($d = 0.28$, $p = .021$), indicating a small-to-moderate effect. First-year students expressed higher levels of societal concern about GenAI compared to upper-year students ($d = 0.13$, $p = .036$), another small effect. 

Two group differences showed more pronounced effects: non--English-speaking students perceived greater clarity in institutional GenAI policies than English-speaking students ($d = 0.41$, $p = .001$), and first-generation college students viewed GenAI as more impactful for their future careers compared to continuing-generation students ($d = 0.38$, $p = .019$). These moderate effect sizes warrant attention, as they reflect more substantial differences in student perceptions. These findings suggest meaningful variation in how students from different backgrounds perceive and engage with GenAI.

\begin{table}[ht]
\centering
\caption{Significant Group Differences in GenAI Attitudes}
\label{tab:group_diffs}
\small
\setlength{\tabcolsep}{4pt}
\begin{tabular}{p{6.5cm} c c}
\toprule
Group Comparison and Finding & $d$ & $p$ \\
\midrule
Non-English speakers found GenAI policy clearer & 0.41 & .001 \\
First-Gens saw greater impact of GenAI on career & 0.38 & .019 \\
Non-English speakers are more accepting of GenAI in writing & 0.28 & .021 \\
Males are more accepting of GenAI in writing & 0.16 & .012 \\
First-years are more concerned about GenAI risks & 0.13 & .036 \\

\bottomrule
\end{tabular}
\vspace{2mm}
\begin{minipage}{0.9\linewidth}
\footnotesize \textit{Note.} $d$ = Cohen’s \textit{d}, a standardized measure of effect size. $p$ = statistical significance value from Welch’s \textit{t}-test.
\end{minipage}
\end{table}

\section{Discussion}

This study provides a psychometrically grounded analysis of undergraduate students' attitudes toward GenAI in academic contexts. Drawing from a structured survey administered at a mid-sized U.S. research university, we identified four latent attitudinal dimensions through EFA and constructed composite scores reflecting the perceived acceptability of GenAI in writing and coursework tasks. Our findings offer a nuanced understanding of student perceptions related to institutional policy clarity, fairness and trust, educational and career implications, and broader societal concerns surrounding GenAI.

First, consistent with prior research, our results reveal a lack of clarity in institutional guidance around GenAI use \cite{johnston2024student}. Many students reported uncertainty about what is permissible, with mixed agreement on whether instructors and institutions have communicated expectations clearly. To reduce confusion and promote ethical GenAI use in academic contexts, institutions need to prioritize transparent and accessible communication of relevant policies.

Second, student attitudes reflected a blend of cautious optimism and critical awareness. While students acknowledged the educational and career potential of GenAI tools, they simultaneously voiced concerns about issues such as fairness, surveillance, data privacy, and biased outputs. This aligns with previous findings that student trust in GenAI is contingent on context, instructor framing, and institutional policy.

Third, task-specific acceptability data demonstrate that students differentiate between low- and high-stakes uses of GenAI. Acceptability was highest for supportive functions (e.g., brainstorming, debugging) and lower for tasks tied directly to graded outputs (e.g., generating full paragraphs). This suggests that students are applying ethical discernment about when GenAI use is appropriate, reinforcing the importance of clearly articulating use cases and academic boundaries.

Lastly, the analysis showed that students’ attitudes and use of GenAI varied across demographic groups. In particular, moderate differences associated with language background and first-generation status suggest a need for further research to understand the factors driving these disparities. Future studies could investigate how students’ educational experiences, access to support resources, or exposure to institutional messaging influence their attitudes toward GenAI. A deeper understanding of these dynamics would support the development of inclusive and responsive policies for AI integration in higher education.

Taken together, the findings emphasize the importance of context-sensitive policy development. Institutions aiming to integrate GenAI tools into learning environments must consider not only the functional affordances of these technologies but also the diverse perspectives students bring to its use. Tailoring communication and guidance to reflect students' backgrounds will be critical to ensuring equitable and ethical GenAI adoption.

\subsection{Limitations and Future Directions}

This study has several limitations. First, the sample was drawn from a single research-intensive U.S. university, which may limit the generalizability of findings to other institutional types or international settings. Future research should incorporate multi-institutional or nationally representative samples to assess whether these patterns extend across varied contexts.

Second, while the EFA supported a meaningful and interpretable factor structure, the internal consistency of the fairness and trust subscale was somewhat lower than ideal. This may reflect the multifaceted nature of the construct or ambiguity in item wording. Further refinement of this subscale—through clearer conceptual definitions and more focused item development—is warranted to improve psychometric reliability.

Lastly, although the survey captured key attitudinal dimensions, qualitative methods such as interviews or focus groups could offer deeper insight into students’ reasoning and lived experiences with GenAI. Future mixed-methods studies could enrich the interpretation of subgroup differences and inform the design of more responsive GenAI education policies.

\section{Conclusion}

As GenAI continues to reshape higher education, understanding how students perceive and engage with these tools is critical to guiding ethical, inclusive, and effective policy development. This study contributes to that goal by offering a psychometrically grounded analysis of undergraduate attitudes toward GenAI, revealing both shared concerns and meaningful subgroup differences.

Our findings highlight the importance of transparent institutional guidance, ethical discernment in student use, and the influence of demographic background on GenAI perceptions. Students are not passive users of new technologies—they actively weigh the risks and benefits based on their academic experiences and personal contexts.

To foster responsible integration of GenAI in education, institutions must go beyond one-size-fits-all approaches. Policies should be clearly communicated, context-sensitive, and attentive to the diverse needs of the student population. As GenAI tools evolve, ongoing research will be essential to monitor shifting attitudes, support equitable adoption, and ensure that these technologies enhance rather than compromise the educational mission.


\section*{Appendix A: Survey Instrument}

\subsection*{Demographics}
\begin{enumerate}[label=\arabic*.]
    \item \textbf{How do you describe your gender identity?}
    \begin{itemize}
        \item Woman
        \item Man
        \item Gender diverse
        \item Non-binary
        \item My gender identity isn't listed. I identify as \_\_\_\_\_\_\_\_\_
        \item Prefer not to say
    \end{itemize}

    \item \textbf{What is your enrolment type?}
    \begin{itemize}
        \item Domestic student
        \item International student
    \end{itemize}

    \item \textbf{Which of the below most closely matches your main area of study?}
    
    (Choose one area if enrolled in a single degree; choose up to two if in a double degree.)
    \begin{itemize}
        \item Humanities/Social Sciences
        \item Architecture/Fine Arts
        \item STEM
        \item Business
    \end{itemize}

    \item \textbf{How many years have you been enrolled in your degree? (Full-time equivalent)}
    \begin{itemize}
        \item Up to 1 year
        \item 2 years
        \item 3 years
        \item 4 years
        \item 5 or more years
    \end{itemize}

    \item \textbf{What language do you mainly speak at home?}
    \begin{itemize}
        \item English
        \item Language other than English
    \end{itemize}

    \item \textbf{Are you the first person in your family to study at a university?}
    \begin{itemize}
        \item Yes
        \item No
        \item Prefer not to say / Don’t know
    \end{itemize}

    \item \textbf{Do you identify as having a disability, health or mental health condition, or as neurodivergent?}
    \begin{itemize}
        \item Yes
        \item No
        \item Prefer not to say
    \end{itemize}
\end{enumerate}

\vspace{1em}
\noindent All attitudinal and behavioral items below were rated on a 5-point Likert scale unless otherwise noted.

\subsection*{Group 1: Perceptions of AI and Institutional Policies (Understanding \& Concerns)}

\textit{Rate the degree to which you agree or disagree with the following statements:}

\begin{itemize}
    \item G1\_1: I feel comfortable approaching my instructor about the use of GenAI.
    \item G1\_2: I am clear on the university’s guidelines for GenAI and academic integrity.
    \item G1\_3: I am clear on how my instructors will respond to suspected inappropriate use of GenAI in the classroom.
    \item G1\_4: I am clear with my own stance on GenAI and academic integrity.
\end{itemize}

\subsection*{Group 2: Concerns About Fairness and Trust}

\begin{itemize}
    \item G2\_1: I am concerned that my professors will not trust me or my work.
    \item G2\_2: I am concerned students who use GenAI to complete assignments will have an unfair advantage.
    \item G2\_3: I am concerned I will receive less authentic feedback from instructors who use GenAI for assessment.
    \item G2\_4: GenAI will disadvantage certain students over others.
\end{itemize}

\subsection*{Group 3: GenAI’s Influence on Education and Career Plans}

\begin{itemize}
    \item G3\_1: GenAI has increased the importance/value of higher education.
    \item G3\_2: GenAI has had an impact on my plan of study.
    \item G3\_3: GenAI has had an impact on my future career plans.
\end{itemize}

\subsection*{Group 4: Broader Concerns About AI and Society}

\begin{itemize}
    \item G4\_1: I am concerned with the environmental impact of GenAI.
    \item G4\_2: I am concerned about personal data privacy and intellectual property.
    \item G4\_3: I believe GenAI will lead to decreased critical thinking.
    \item G4\_4: I am concerned about biased outputs in GenAI.
    \item G4\_5: I am concerned about inaccurate outputs in GenAI.
\end{itemize}

\subsection*{Group 5: Acceptability of GenAI in Writing Tasks}

\textit{In terms of upholding academic integrity, how acceptable do you find the use of GenAI in your writing tasks?}

(Response options: Always Unacceptable, Unacceptable in Most Cases, Uncertain, Acceptable in Most Cases, Always Acceptable)

\begin{itemize}
    \item G5\_1: Checking for grammar.
    \item G5\_2: Revising your own written prose for formality/tone/style.
    \item G5\_3: Generating entire sentences or paragraphs from a student-generated prompt.
    \item G5\_4: Seeking feedback on student-generated writing (without direct revision of that writing).
    \item G5\_5: Finding sources (background research and locating sources).
    \item G5\_6: Brainstorming ideas.
\end{itemize}

\subsection*{Group 6: Acceptability of GenAI in Coursework}

\textit{Assuming your professor has not prohibited it, how acceptable do you find the use of GenAI in your coursework?}

(Response options: Always Unacceptable, Unacceptable in Most Cases, Uncertain, Acceptable in Most Cases, Always Acceptable)

\begin{itemize}
    \item G6\_1: Generating sample/practice problems.
    \item G6\_2: Generating study guides.
    \item G6\_3: Generating creative content.
    \item G6\_4: Supporting coding tasks / fixing coding errors.
    \item G6\_5: Supporting / checking language translations.
\end{itemize}

\section*{Appendix B}
\begin{figure}[ht]
\centering
\includegraphics[width=0.4\textwidth]{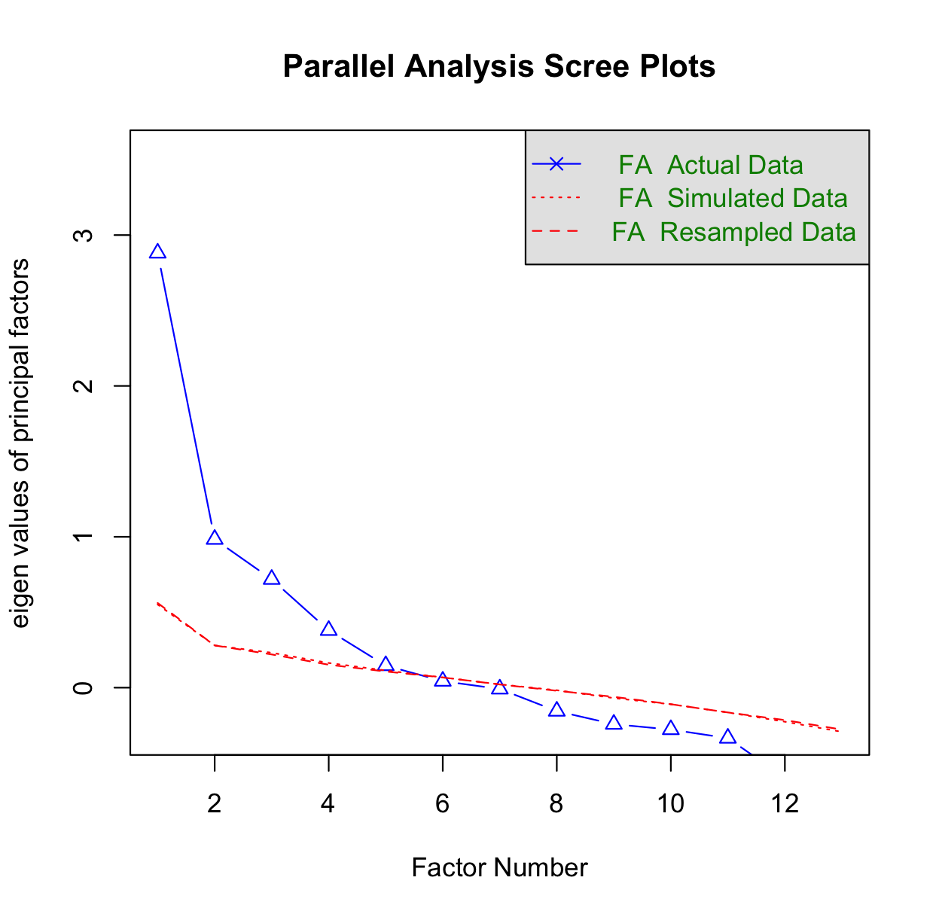}
\caption{Scree plot from parallel analysis}
\label{fig:scree}
\end{figure}

\end{document}